\documentclass[prd,aps,showpacs,preprintnumbers,amssymb]{revtex4}
\usepackage{axodraw}
\usepackage{color}
\usepackage{epsf}

\def\e3p{$\eta \rightarrow 3 \pi$}

\begin{document}
\title{%
\hfill{\normalsize\vbox{%
\hbox{}
 }}\\
{  How important is  $i\epsilon$ in QFT?}}

\author{Amir H. Fariborz
$^{\it \bf a}$~\footnote[1]{Email:
 fariboa@sunyit.edu}}
\author{Renata Jora
$^{\it \bf b}$~\footnote[2]{Email:
 rjora@theory.nipne.ro}}

\author{Joseph Schechter
 $^{\it \bf c}$~\footnote[4]{Email:
 schechte@phy.syr.edu}}

\affiliation{$^{\bf \it a}$ Department of Mathematics/Physics, State University of New York, Polytechnic Institute,  Utica, NY 13504-3050, USA}
\affiliation{$^{\bf \it b}$ National Institute of Physics and Nuclear Engineering PO Box MG-6, Bucharest-Magurele, Romania}

\affiliation{$^ {\bf \it c}$ Department of Physics,
 Syracuse University, Syracuse, NY 13244-1130, USA}

\date{\today}

\begin{abstract}
We discuss the role of  $i\epsilon$ in quantum field theories and suggest that it can be identified  with the dimensional regularization parameter $i\epsilon=4-d$  thus clarifying and simplifying issues related to the infrared divergences without altering any of the present knowledge in QFT. We further present the relevance of this assumption for  the optical theorem.
\end{abstract}
\maketitle

\section{Introduction}

Consider the path integral approach to quantum mechanics for a theory with the Hamiltonian $H$ \cite{Srednicki}.
We denote by $|n\rangle$ the eigenstate of the Hamiltonian $H$ with the eigenvalue $E_n$.  Then $|0\rangle$ and $E_0$ correspond to the ground state. If $|q\rangle$ is the eigenstate
corresponding to the momentum operator $q$ one can write:
\begin{eqnarray}
&&|q,t\rangle=e^{iHt}|q\rangle=\sum_{n=0}^{\infty}e^{iHt}|n\rangle\langle n |q\rangle=
\sum_{n=0}^{\infty}\Phi_n(q)e^{iE_nt}|n\rangle,
\label{rez324567}
\end{eqnarray}
where $\Phi_n(q)$ is the wave function for the energy $E_n$. In order to extract the ground state we make the Feynman replacement $H\rightarrow (1-i\epsilon)H$  or equivalently $t\rightarrow (1-i\epsilon)t$. Then in the limit $t\rightarrow -\infty$  with $\epsilon$ held fixed one obtains that the final state in Eq. (\ref{rez324567}) is the ground state.

If we go further to the path integral formalism (assume a simple scalar theory) the substitution $t\rightarrow t(1-i\epsilon)$ is equivalent to $k^0\rightarrow k^0(1+i\epsilon)$ and subsequently  $(k^2-m^2)\rightarrow (k^2-m^2+i\epsilon)$ \cite{Peskin}. Note here that in standard quantum field theories $i\epsilon$ is a quantity with hidden $mass^2$ dimension because one makes the identification $2k_0^2\epsilon \approx \epsilon$. Thus the partition function for a real scalar field theory in the Fourier space is given by:
\begin{eqnarray}
Z=\prod_n d{\rm Re}\Phi(p_n) d{\rm Im}\Phi(p_n)\exp[-\frac{i}{V}\sum_n(m^2-k^2-i\epsilon)[({\rm Re}\Phi(p_n))^2+({\rm Im}\Phi(p_n))^2]+...].
\label{part4567}
\end{eqnarray}

Since $\exp[-\epsilon[({\rm Re}\Phi(p_n))^2+({\rm Im}\Phi(p_n))^2]]\rightarrow 0$ as ${\rm Re(Im)}\Phi(p_n)\rightarrow\pm\infty$  with $\epsilon$ held fixed the presence of the $i\epsilon$ term ensures the convergence of the gaussian integrals in the path integral formalism. The $i\epsilon$ term also plays an important role in the direction of analytic continuation in the conversion from the Minkowski space to the Euclidean one in the evaluation of the Feynman integrals.

It is important to note that although infinitesimal the parameter $\epsilon$ is held fixed when the momenta $k_{\mu}^2$ (without summation over $\mu$) go to zero. This can be seen from the mere definition or by observing that $\epsilon$ is held fixed when ${\rm Re}\Phi(p_n)\rightarrow \infty$. But,
\begin{eqnarray}
{\rm Re}\Phi(p_n)\approx\int d^4 x \Phi(x)\cos(p_n x)
\label{ev5467}
\end{eqnarray}
is maximum so it attains the infinite limit when $p_n\rightarrow 0$.

In conclusion in the path integral formalism one keeps $\epsilon$ fixed when evaluating or taking some limits of momenta. But we know another small (infinitesimal) parameter with similar properties. This is the infrared regulator  $\mu^2$ for theories with infrared divergences as many of the known ones are.
A typical Feynman integral for a theory with infrared divergences has the expression in the  Minkowski space:
\begin{eqnarray}
\int \frac{d^4 k}{(2\pi)^4}\frac{1}{(k^2-\mu^2+i\epsilon)^n},
\label{fey5654}
\end{eqnarray}
where $\mu$ is a small infrared regulator.

First one performs a contour integral in the $k^0$ plane where the poles are at \cite{Weinberg}:
\begin{eqnarray}
k^0=\pm \sqrt{|\vec{k}|^2+\mu^2-i\epsilon}=\pm\sqrt{(|\vec{k}|^2+\mu^2)^2+\epsilon^2}\exp[-i\frac{1}{2}\arctan\frac{\epsilon}{|\vec{k}|^2+\mu^2}].
\label{dezv54678}
\end{eqnarray}
In order to be able to rotate the contour integral to the imaginary axis one needs to have the positive pole (on the real axis) below the imaginary axis and the negative one above the imaginary axis. The procedure must be performed also for $|\vec{k}|^2=0$ case in which two situations are possible. If we hold $\epsilon$ fixed and take the limit $\mu^2\rightarrow 0$ the poles are located on the imaginary axes  and the conversion cannot be made. If we keep $\mu^2$ fixed and make $\epsilon\rightarrow 0$ the poles are located on the real axis and the rotation cannot be performed.

\section{The possible multiple roles of $i\epsilon$}

From our previous discussion one can deduce that $i\epsilon$ and $\mu^2$ have similar properties from the point of view of their convergence to zero in the path integral formalism.
On the other hand it is known that one can introduce a small mass as an infrared regulator or in the dimensional regularization mechanism \cite{Carlos}, \cite{Hooft} the infrared regulator is related to the
dimension $d=4-\epsilon$ such that $\epsilon_{IR}=-\epsilon_{UV}$ \cite{Schwartz}. Thus  in the end one needs two dimensions in the dimensional regularization approach: one $d>4$ in the ultraviolet and one $d<4$ in the infrared. An obvious question then appears. Can we fix at least a part of the problem by introducing not two or three infinitesimal parameters but a single one with multiple functions? If so then this parameter must be related somehow to dimensional regularization.

Let us consider the kinetic term of a real scalar field in the Fourier space:
\begin{eqnarray}
{\cal L}=\int \frac{d^d p}{(2\pi)^d}\Phi(p)(p^2-m^2)\Phi(-p).
\label{four6578}
\end{eqnarray}
We can rewrite this in spherical coordinate as:
\begin{eqnarray}
{\cal L}= \int \frac{d p^d}{(2\pi)^d}d \Omega_d \Phi(p)(p^2-m^2)\Phi(-p).
\label{four8979}
\end{eqnarray}
Here $\Omega_d$ are the angular coordinates in the $d$ dimensional space. We shall further assume that $d=4-i\epsilon$ is complex. We perform the change of variables:
\begin{eqnarray}
&&p^d=q^4
\nonumber\\
&&p^2=q^{2+i\epsilon/2}=q^2(1+i\epsilon/2\ln(q)),
\label{cha56748}
\end{eqnarray}
where we expand in the small parameter $\epsilon$.
We can make abstraction of the overall factors proportional to $i\epsilon$ multiplying the kinetic term and write:
\begin{eqnarray}
{\cal L}={\rm const} \int d q q^3\frac{1}{(2\pi)^4}\Phi(q)(q^2+ i\frac{\epsilon}{2}q^2\ln(q)-m^2)\Phi(-q).
\label{rez2121}
\end{eqnarray}

Assuming $\epsilon$ small ($\frac{1}{2}\epsilon q^2 \ln(q)=\epsilon$) the Lagrangian in the coordinate space can as well be written as:
\begin{eqnarray}
{\cal L}=-\int d^4 x \Phi(x)(\partial^2+m^2-i\epsilon)\Phi(x).
\label{rez4356712}
\end{eqnarray}

 It may seem that by making this expansion from the beginning one may spoil some of the nice features of dimensional regularization.  The point is that the usual $d$ dimensional integrals are maintained in the interaction terms such that when one calculates corrections the standard $d$ dimensional Feynman integrals appear. Thus the above expansion does not alter in any way the dimensional regularization scheme but only leads to the correct form of the propagators.

 Since a fractional dimension real or complex still does not make sense we claim that there is no significant difference with the standard dimensional regularization formalism with the exception that the small parameter $\epsilon$ in the dimensional regularization is replaced with $i\epsilon$.  This ensures the correct convergence of the gaussian integrals in the partition function and also provide the direction of analytic continuation when one makes the conversion to the euclidean space.  So the standard procedure should work as well.
 Moreover the same $i\epsilon$ should play a role in the regularization of infrared divergences and thus work for both the infrared and ultraviolet regimes. We thus replaced three parameters by a single one that performs all the functions of the three ones.

Let us show how the new small parameter works. For that consider a typical Feynman integral with infrared and ultraviolet divergences:
\begin{eqnarray}
\int d^d k\frac{1}{(2\pi)^d}\frac{1}{(k^2+i\epsilon)^2}=\frac{1}{8\pi^2i\epsilon}-\frac{\ln{(-i\epsilon)}}{16\pi^2}+...\Rightarrow\frac{1}{8\pi^2i\epsilon}.
\label{newrs3456}
\end{eqnarray}

Since the same $\epsilon$  appears in the term $\frac{1}{\epsilon}$ and $\ln\epsilon$ and the limit $\lim_{\epsilon \rightarrow 0}\epsilon\ln\epsilon=0$ one can include the  logarithmic divergence in the first term in the r. h. s. of Eq. (\ref{newrs3456}) and thus obtain the regular divergence in the dimensional regularization scheme valid for both the infrared and ultraviolet regions.

In  a more phenomenological context  let us consider the vacuum polarization amplitude for QED:
\begin{eqnarray}
\Pi^{\mu\nu}_2(q^2)=(q^2g^{\mu\nu}-q^{\mu}q^{\nu})\Pi_2(q^2),
\label{rez32455}
\end{eqnarray}
where at one loop,
\begin{eqnarray}
\Pi_2(q^2)=\frac{-2\alpha}{\pi}\int_0^1 dx x(1-x)[\frac{2}{i\epsilon}-\ln[m^2-i\epsilon-x(1-x)q^2]-\gamma+\ln[4\pi]].
\label{rz32456}
\end{eqnarray}
In the limit $m^2=0$ one obtains:
\begin{eqnarray}
&&\Pi_2(q^2)=\frac{-2\alpha}{\pi}\int_0^1 dx  x(1-x)[\frac{2}{i\epsilon}-\ln[x(1-x)q^2]-\gamma+\ln[4\pi]]
\nonumber\\
&&\Pi_2(0)=\frac{-2\alpha}{\pi}\int_0^1 dx x (1-x)[\frac{2}{i\epsilon}-\ln[-i\epsilon]-\gamma+\ln[4\pi]].
\label{rez43567}
\end{eqnarray}
If one includes the logarithmic term in the ultraviolet divergence,
\begin{eqnarray}
\frac{2}{i\epsilon}-\ln[-i\epsilon]=\frac{2}{i\epsilon}-\ln[-i\epsilon/M^2]-\ln[M^2]=\frac{2}{i\epsilon}-\ln[M^2],
\label{rez435678}
\end{eqnarray}
where $M^2$ is an arbitrary scale, then one obtains:
\begin{eqnarray}
\Pi_2(q^2)=\frac{\alpha}{3\pi}[\ln[-q^2/M^2]-5/3+...],
\label{rezxcv}
\end{eqnarray}
a known result that makes perfect sense even in the limit $m=0$.

Thus the identification of the $i\epsilon$ parameter (up to a finite scale) with the dimensional regularization term $i\epsilon=4-d$ can help dealing with the infrared divergences without the usual complications.
\section{Relation to the optical theorem and discussion}

We consider again a scalar field theory. The all orders propagator is given by:
\begin{eqnarray}
{\rm Propagator}(p^2)=\frac{i}{p^2-m^2+M(p^2)}.
\label{req324}
\end{eqnarray}
This propagator will receive an imaginary part according to the optical theorem. Upon the use of the renormalization conditions $M(m^2)=0$ and $M'(m^2)=0$ for $p^2$ close to $m^2$ the propagator will become:
\begin{eqnarray}
{\rm Propagator}(p^2)=\frac{i}{p^2-m^2+{\rm Im}M(p^2).}
\label{op8er79}
\end{eqnarray}

Then the optical theorem claims that for a narrow decay width in the vicinity of $p^2\approx m^2$ the propagator can be written as:
\begin{eqnarray}
{\rm Propagator}(p^2)=\frac{i}{p^2-m^2+i m \Gamma},
\label{op879}
\end{eqnarray}
where $\Gamma$ is the full decay width of the corresponding scalar particle.

We claim that there is a direct connection between the Eq. (\ref{op879}) and the all orders formula for the propagator as resulted from  Eq. (\ref{rez4356712}):
\begin{eqnarray}
{\rm Propagator}(p^2)=\frac{i}{p^2-m^2+i\epsilon}.
\label{op87459}
\end{eqnarray}

Let us justify this. Assume we compute an all order propagator in the standard  approach  for a scalar theory and we get the equations (\ref{op8er79}) and (\ref{op879}). Then for the same theory we work from the beginning in the euclidean space.  There are no negative arguments of the logarithms and the corrections to the propagator are all real.  After renormalization and  the conversion to the Minkowski space the propagator will have exactly the expression in Eq. (\ref{op87459}).  Thus we can assume that the following relation holds:
\begin{eqnarray}
\epsilon=m\Gamma.
\label{res43567}
\end{eqnarray}
This relation says even more for the situation described in Eq. (\ref{rez2121}) where we retrieve the original form for the $i\epsilon$ term as $i\epsilon/2p^2\ln(p)$. Moreover from the standard dimensional and cut-off regularization procedures one can write the equivalence:
\begin{eqnarray}
\frac{2}{\epsilon}=\ln(\Lambda^2).
\label{equiv4356}
\end{eqnarray}

Since the logarithms are not well defined we simply introduce an arbitrary scale (which may be considered the renormalization scale) to make the arguments dimensionless. Then we just write for $p^2=m^2$:
\begin{eqnarray}
&&m\Gamma=\frac{\epsilon}{2}m^2\ln(m/M)
\nonumber\\
&&\frac{\epsilon}{2}=\frac{1}{\ln(\Lambda^2/M^2)}.
\label{re90z43567}
\end{eqnarray}
Here one employs the natural expression obtained in this work for the $\epsilon$ term as opposed to Eq. (\ref{op87459}) where $i\epsilon$ is just the standard  notation for the imaginary part of the denominator.
First note that the first relation in Eq. (\ref{re90z43567}) is only approximate as may receive corrections and depends on how narrow is the decay width. Second it will receive a minus sign for a gauge boson because of the inverse sign of the propagator and also a factor of $1/2$ for the fermions from applying Eq. (\ref{four8979}) to the fermion kinetic term. Thus one obtains for gauge bosons and fermions:

\begin{eqnarray}
&&m\Gamma_A=-\epsilon/2m^2\ln(m/M)
\nonumber\\
&&m\Gamma_f=\epsilon/4 m^2\ln(m/M).
\label{re434343}
\end{eqnarray}

\begin{figure}
\begin{center}
\epsfxsize = 8cm
 \epsfbox{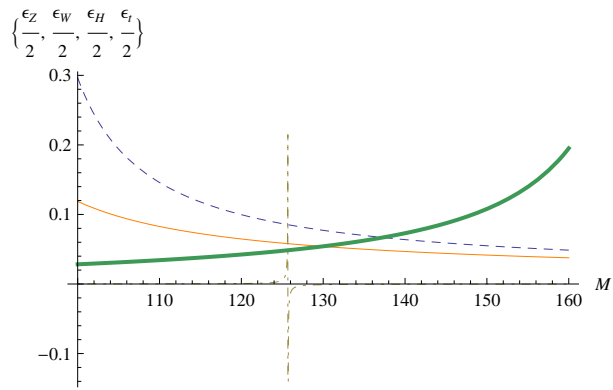}
\end{center}
\caption[]{%
Plots of $\epsilon_Z/2$ (dashed line), $\epsilon_W/2$ (orange line), $\epsilon_H/2$ ( dotdashed line) and $\epsilon_t/2$ (thick line) as functions of the renormalization scale $M$. The region of intersection for the four $\epsilon$'s  indicates the range of values for the natural cut-off scale of the theory $\Lambda$.
}
\label{cy65}
\end{figure}

We shall apply the Eqs. (\ref{re90z43567}) and (\ref{re434343}) to the W, Z, Higgs boson and the top quark of the standard model to get:
\begin{eqnarray}
&&\Gamma_Z/m_Z=-\epsilon_Z/2\ln(m_Z/M)
\nonumber\\
&&\Gamma_W/m_W=-\epsilon_W/2\ln(m_W/M)
\nonumber\\
&&\Gamma_H/m_H=\epsilon_H/2\ln(m_H/M)
\nonumber\\
&&\Gamma_t/m_t=\epsilon_t/4\ln(m_t/M).
\label{rel789}
\end{eqnarray}
We shall use the central values for the masses and decay widths as taken from \cite{PDG}: $m_Z=91.1876$ GeV, $\Gamma_Z=2.4952$ GeV; $m_W=80.385$ GeV, $\Gamma_W=2.085$ GeV; $m_H=125.7$ GeV, $\Gamma_H=4.07\times 10^{-3}$ GeV; $m_t=173.3$ GeV, $\Gamma_t=1.35$ GeV. We plot the quantities $\epsilon_Z/2$, $\epsilon_W/2$, $\epsilon_H/2$ and $\epsilon_t/2$ as functions of the renormalization scale for a range of $100-160$ GeV for M. The reason that we chose such a small interval is that only in this range of values a possible intersection of the corresponding graphs may occur. The results are depicted in Fig. \ref{cy65}.

As expected we find a region of intersection such that for a fixed M close to the mass of the Higgs boson the parameter $\frac{2}{\epsilon}$  is in the range $0.048\leq\frac{2}{\epsilon}\leq 0.085$ which would correspond to a cut-off scale in the range $0.043\times 10^6\leq \Lambda \leq 3.82 \times 10^6$ GeV. If M is allowed to vary in the range $125.6-137.3$ Gev the interval for the cut-off becomes $0.242\times 10^6\leq \Lambda \leq 3.82 \times 10^6$ GeV. This results may improve if one takes into account the running of the masses with the scales.

 In conclusion the $i\epsilon$ parameter (up to a finite scale) can be identified with the dimensional regularization parameter $i\epsilon=4-d$ and thus play multiple roles. On the other hand we expect that all known advantages of the dimensional regularization scheme are maintained as there is no difference in the standard procedure if the parameter $\epsilon$ is taken real or imaginary. In the end one can always make again $\epsilon$ real. However our assumption replaces three parameters by a single one and make the whole dimensional regularization approach simpler and more feasible.

Moreover the relation between the $i\epsilon$ term in the Lagrangian and the dimensional regularization parameter can be very fruitful in applications of the optical theorem and can give information with regard to the imaginary part of the propagator and the decay width for the case when the Breit Wigner formula holds.

\section*{Acknowledgments} \vskip -.5cm

The work of R. J. was supported by a grant of the Ministry of National Education, CNCS-UEFISCDI, project number PN-II-ID-PCE-2012-4-0078.

\end{document}